\documentclass[aps,prl,twocolumn,superscriptaddress,floatfix,footinbib]{revtex4-2}

\usepackage{hyperref}
\usepackage{natbib}
\usepackage{etoolbox}
\apptocmd{\thebibliography}{\raggedright}{}{}

\usepackage[utf8]{inputenc}
\usepackage{subfiles}
\usepackage{amsfonts}
\usepackage{graphicx}
\usepackage{physics}
\usepackage{float}
\usepackage{amsmath}
\usepackage{mathtools}
\usepackage[english]{babel}
\usepackage{csquotes}

\makeatletter
\newcommand{\prlsection}[1]{%
  \par\vspace{1ex}%
  \textit{#1.---}\ignorespaces
}
\makeatother

\usepackage{tcolorbox}

\newcommand{\U}{\mathcal{U}} 
\newcommand{\PF}{\mathcal{P}} 
\newcommand{\M}{\mathcal{M}} 
\newcommand{\C}{\mathcal{C}} 

\newcommand{\hrho}{\hat{\rho}} 
\newcommand{\hkr}{\hat{\kappa}} 

\newcommand{\sbraket}[2]{\left(#1|#2\right)}
\newcommand{\ssbraket}[2]{(#1|#2)}

\newcommand{\bnt}[2]{\beta_{#1\,#2}}
\newcommand{\atn}[2]{\alpha_{#1\,#2}}

\newcommand{\uba}{Universidad de Buenos Aires, Facultad de Ciencias Exactas y Naturales, Departamento de Física. Buenos Aires 1428, Argentina}
\newcommand{\ifiba}{CONICET - Universidad de Buenos Aires, Instituto de Física de Buenos Aires (IFIBA). Buenos Aires 1428, Argentina}

\newcommand{\thetitle}{Classical chaotic signatures in Krylov space}

\begin{document}

\title{\thetitle}
    \author{Gast\'on F. Scialchi}
    \email[E--mail address: ]{gscialchi@df.uba.ar}
    \affiliation{\uba}
    \affiliation{\ifiba}

    \author{Augusto J. Roncaglia}
    \email[E--mail address: ]{augusto@df.uba.ar}
    \affiliation{\uba}
    \affiliation{\ifiba}

    \author{Diego A. Wisniacki}
    \email[E--mail address: ]{wisniacki@df.uba.ar}
    \affiliation{\uba}
    \affiliation{\ifiba}

    \begin{abstract}
        In recent years, Krylov space approaches have emerged as a useful framework for characterizing dynamical chaos
        in a wide variety of physical systems.
        However, the connection between Krylov dynamics and the established structures of classical chaos remains largely unexplored.
        Classical chaos is characterized by exponential stretching and mixing.
        For quantum maps with a classical limit, we show that
        these two signatures are encoded in different structures within Krylov space:
        stretching is encoded in the geometry of the Krylov operators as determined by the Lyapunov exponent,
        while mixing manifests itself as an exponential decay of the Arnoldi sequences in agreement with the Ruelle-Pollicott resonances.
    \end{abstract}

\maketitle

The two fundamental properties of classical chaos are the exponential separation of trajectories
and their posterior mixing.
Since these properties cannot be translated into the quantum regime,
the field of quantum chaos was mainly concerned with other manifestations of the underlying classically chaotic dynamics.
For example, the relationship between the spectral statistics of a quantum system
and the chaoticity of its underlying classical counterpart~\cite{haake1991quantum, mehta2004random, wimberger2014nonlinear}.
A complementary time-dependent dynamical picture has also been developed,
with tools such as the Loschmidt echo~\cite{Wisniacki2012, PhysRevLett.86.2490},
a probe of time-reversibility under perturbations,
and the out-of-time-order correlator (OTOC)~\cite{shenker2015stringy, swingle2018unscrambling,Garcia-Mata:2023,PhysRevLett.121.210601},
whose growth is associated with the spread of quantum information.
These dynamical quantities reflect chaotic properties of the underlying classical system,
if it exists.
Where it does not, chaoticity in quantum systems
is often defined according to how the aforementioned quantities behave.

Krylov space methods have emerged as a powerful tool to study quantum dynamics and chaos~\cite{NANDY20251, rabinovici2025krylovcomplexity}.
They consist of an iterative orthogonal expansion~\cite{viswanath2008recursion} for the evolution of an observable or state,
mapping it to that of a wavefunction on a one-dimensional chain
where each site represents an element of the Krylov basis.
The wavefunction is initially localized on one end of the chain
and ventures further into Krylov space as time evolves,
inspiring notions of evolution complexity based on its spreading~\cite{Parker2019}.
They entered the dictionary of quantum chaos through the ``universal operator growth hypothesis"~\cite{Parker2019},
which states that under chaotic Hamiltonian dynamics
the chain's hopping amplitudes (given by the Lanczos coefficients)
are such that it supports maximal complexity growth.
A complementary picture has been proposed
in general unitary quantum dynamics, where an underlying static Hamiltonian may not exist:
the hopping amplitudes (now determined by the Arnoldi sequences)
give rise to a ``maximally ergodic regime"
where maximal complexity growth~\cite{PhysRevB.111.014309} is attained in chaotic systems.

An extensive literature soon developed~\cite{Rabinovici2021, Rabinovici_2022_localization, Rabinovici_2022_integrability}
relating the structure and dynamics of Krylov space to quantum chaos across a wide variety of settings,
ranging from many-body spin chains~\cite{PhysRevE.107.024217, PhysRevE.109.054209, PhysRevE.110.034201, PhysRevE.111.014220}
to quantum field theories, holographic systems and gravitational models~\cite{spcr-jgm6, Bhattacharjee_2023, rabinovici2023bulk, rabinovici2025krylovcomplexity},
and beyond~\cite{NANDY20251}.
However, comparatively little has been established on the manifestations of classical chaos in Krylov space.

Classical chaos is characterized by two complementary dynamical properties.
The first one is the exponential separation of nearby trajectories,
quantified by the Lyapunov exponent and manifested as the stretching of distributions in phase space.
A stronger manifestation is mixing which,
as a consequence of the progressive stretching and folding in a compact phase space,
produces the relaxation of phase space distributions and the decay of correlation functions.
In strongly chaotic systems, this decay is exponential and determined by the Ruelle-Pollicott resonances~\cite{Ruelle1986, 10.4310/jdg/1214440727, Pollicott1985}.
In Krylov space,
ergodicity is associated with the efficient transport of the Krylov wavefunction along the chain~\cite{PhysRevB.111.014309},
whose hopping amplitudes are determined by the Arnoldi sequences in unitary dynamics.
Thus, the question is raised as to whether the classical mechanisms underlying mixing are also encoded in the structure of the Krylov chain.

In this Letter we show, in quantized classical maps,
that the Lyapunov exponent can be found in the geometry of the Krylov operators,
which we prove analytically for a simple model of quantum chaos.
On the other hand,
we find the Ruelle-Pollicott resonances manifested in the exponential decay of the Arnoldi sequences,
showing they play an important role in the approach to the maximally ergodic regime in Krylov space.

We study quantized classical area-preserving maps on the torus.
These are among the simplest systems with both an exponential separation of trajectories and a compact phase space.
Both their Lyapunov exponents and Ruelle-Pollicott resonances can be computed,
and their quantum versions are described by simple unitary Floquet operators.
Here we consider the perturbed Arnold's cat map~\cite{arnold1989ergodic, PhysRevLett.121.210601, PhysRevE.62.1977} on the unit torus, which is defined by
\begin{equation}
    \begin{split}
        p' &= p + q -  2\pi k \sin({2\pi q}) \mod 1\\
        q' &= q + p' + 2\pi k \sin({2\pi p'}) \mod 1
    \label{eq:c_cat_map}.
    \end{split}
\end{equation}
The perturbation is introduced to avoid non-generic behavior in the linear map,
and it stays close to its $k=0$ Lyapunov value $\Lambda = \ln[(3+\sqrt{5})/2]$
for small perturbations.
Upon quantization,
the system's periodicity in phase space in both $q$ and $p$ directions (torus structure)
leads to a Hilbert space of finite dimension $N$~\cite{P_Leboeuf_1990} requiring the consistency relation $2\pi\hbar N = 1$,
such that the classical limit is obtained by taking $N\to\infty$.
This discretizes the phase space in the sense that the $\hat{q}$ and $\hat{p}$ operators have a finite number of eigenstates.
The resulting quantum evolution is described by the Floquet unitary
\begin{equation}
    \hat{U} =
    e^{-i 2\pi N( \frac{1}{2} \hat{p}^2 - k\cos(2\pi \hat{p}))}
    e^{i 2\pi N( \frac{1}{2} \hat{q}^2 + k\cos(2\pi \hat{q}))}
    \label{eq:q_cat_map}.
\end{equation}

The evolution of a phase space density $\rho(x)$ generated by a classical area-preserving map $x' = \M(x)$
can be written as
\begin{equation}
    \rho'(x) = \PF \rho(x) = \int dx' \, \delta(x - \M(x')) \rho(x')
    \label{eq:perron_frobenius},
\end{equation}
where $\PF$ is the Perron-Frobenius (PF) evolution operator.
This operator is unitary when acting on the space of square-integrable functions $L^2$,
but not when considering an extended space allowing for distributions.
In uniformly hyperbolic chaotic systems its spectrum consists of an essential and a point spectrum~\cite{reed1981functional},
the latter of which defines the Ruelle-Pollicott resonances $\{\lambda_i\}$.
The largest resonance $\lambda_0 = 1$ corresponds to the invariant density
and the rest determine the decay of correlation functions and relaxation,
which for a smooth $\rho(x)$ will be dominated by the leading nontrivial resonance $\lambda_1$ in the absence of degeneracies.
However, most systems are not uniformly hyperbolic but have a mixed phase space,
supporting both regular and chaotic dynamics.
In this case, unitarity is relaxed by systematically introducing coarse-graining and truncation schemes
such that the stable eigenvalues closest to the unit circle can be identified as generalized resonances~\cite{Christopher_Manderfeld_2001, Joachim_Weber_2001, PhysRevE.62.4769, PhysRevE.62.1977}.

Krylov space is obtained from an iterative orthogonal expansion of the evolution.
Given the evolution of an observable $\hrho_0$ generated by a unitary superoperator $\hrho_{t+1} = \U \hrho_t = \hat{U} \hrho_t {\hat{U}^\dagger}$,
one obtains the Krylov basis $\{\hkr_0, \hkr_1, ...\}$ by defining an operator inner product $\ssbraket{\hat{A}}{\hat{B}}$ and orthonormalizing the sequence $\{\hrho_0, \hrho_1, ...\}$.
Any one of several procedures may be used to that end, such as the Gram-Schmidt process, Szegő recurrence or Arnoldi iteration.
The latter is defined by
\begin{equation}
    b_n \hkr_n = \U \hkr_{n-1} - \sum_{\nu=0}^{n-1} \sbraket{\hkr_\nu}{\U\hkr_{n-1}} \hkr_\nu
    \label{eq:arnoldi},
\end{equation}
where $\hkr_0 = \hrho_0/\norm{\hrho_0}$.
In the Krylov basis, the superoperator is represented by a matrix with an upper Hessenberg form parametrized by the Arnoldi sequences
\begin{align}
    b_n = \sbraket{\hkr_n}{\U\hkr_{n-1}},\,\, &
    a_n = \sbraket{\hkr_n}{\U\hkr_n},\,\,
    c_n = \sbraket{\hkr_0}{\U\hkr_n},
    \label{eq:arnoldi_sequences}
\end{align}
such that the remaining nonzero elements are
$\U_{m n} = \sbraket{\hkr_m}{\U\hkr_n} = c_n a_m/c_m \quad \forall\, m\leq n$.
These sequences are real, with $b_n \geq 0$, and unitarity requires to obey $\abs{\U_{m n}} \leq 1$.
Expanding the evolution in the Krylov basis $\hrho_t = \sum_{n=0}^t \bnt{n}{t}\,\hkr_{n}$
maps it to that of a \textit{wavefunction} on a quasi one-dimensional chain
\begin{equation}
    \bnt{n}{t+1} = b_n \bnt{n-1}{t} + a_n \bnt{n}{t} + \frac{a_n}{c_n} \sum_{\nu>n}^{t} c_\nu \bnt{\nu}{t}
    \label{eq:beta_evo}
\end{equation}
initially localized on its left-most site $\bnt{n}{0} = \delta_{n\,0}$.
Thus, the transport properties of the wavefunction on the chain are determined by the Arnoldi sequences.
Their properties are known to reflect the quantum-chaotic nature of the system~\cite{Rabinovici_2022_localization, PhysRevE.107.024217, PhysRevE.109.054209, PhysRevB.111.014309, PhysRevE.111.014220, PhysRevE.111.L052202}.
In particular, Ref.~\cite{PhysRevB.111.014309} argues that chaotic systems
exhibit a universal \textit{maximally ergodic} regime in Krylov space.
This regime is characterized by memory-less Krylov dynamics,
in which the autocorrelation of each Krylov operator vanishes after a single time step, $\sbraket{\hkr_{n}}{\U\hkr_{n}} = 0$.
Since $\hkr_n = \U \hkr_{n-1}$ is then orthogonal to all previous Krylov operators, it is itself the next one, $\hkr_{n}=U\hkr_{n-1}$.
As a consequence, the Arnoldi sequences satisfy $a_n = 0$, $b_n = 1$ and $c_n = 0$,
which implies maximally efficient ballistic transport of the wavefunction in Eq.~\eqref{eq:beta_evo}.

Here we make use of a Hilbert-Schmidt-like inner product
\begin{equation}
    \ssbraket{\hat{A}}{\hat{B}} = \frac{1}{2\pi\hbar}\Trace(\hat{A}^\dagger \hat{B})
    \label{eq:product}
\end{equation}
and consider smooth observables given by traceless pseudo-density matrices,
which we express via the Glauber-Sudarshan P-representation~\cite{PhysRev.131.2766, PhysRevLett.10.277} as
\begin{equation}
    \hrho_0 = \int dx\,\rho_0(x) \ketbra{\alpha(x)} - \frac{\mathbb{I}}{N}
    \label{eq:p_representation},
\end{equation}
where $\ket{\alpha(x)}$ is the coherent state centered at a point $x$ in phase space.
In particular, we will consider $\hrho_0$
defined by a periodized Gaussian distribution on the unit torus as
\begin{equation}
    \rho_0(q, p) =
    \frac{1}{2\pi\sigma^2} \sum_{n,m\in\mathbb{Z}}
    e^{-\frac{1}{2\sigma^2}\left[(q-q_0-n)^2 + (p-p_0-m)^2\right]}
    \label{eq:rho0_torus}.
\end{equation}
The Krylov space constructed by Eqs.~\eqref{eq:product} and \eqref{eq:p_representation} obeys a quantum-to-classical correspondence~\footnote{Since we are arguing about classical signatures, we define the construction such that a limiting classical Krylov space is well defined. The use of the Glauber-Sudarshan P-representation is not completely necessary for this, one may as well use the Wigner function or Husimi distribution, and the classical limit would be the same~\cite{Scialchi_2026}}:
in the limit $N\to\infty$
the quantum Krylov space becomes equivalent to that defined by the classical evolution of $\rho_0(x)$ [Eq.~\eqref{eq:perron_frobenius}] and the $L^2$ inner product~\cite{Scialchi_2026}.
The subtraction of the identity in Eq.~\eqref{eq:p_representation} makes $\hrho_0$ traceless,
and thus orthogonal to the uniform density.
This is important since one expects chaotic dynamics in a compact phase space to relax the initial condition into a quasi-uniform density,
leading to Krylov dynamics dominated by the uniform density (see the Supplemental Material).

Although the Arnoldi iteration has become the standard procedure to construct the Krylov space of non-Hermitian generators,
for unitary dynamics the Szegő recurrence is particularly convenient.
Indeed, the Arnoldi sequences are not independent:
they can all be constructed from the sequence of Verblunsky coefficients.
The relation between these coefficients and integrable or chaotic dynamics was studied in Ref.~\cite{PhysRevE.111.L052202}.
In the Supplemental Material we review this construction and express the following results in such terms.

\begin{figure}[t]
    \centering
    \includegraphics[width=\linewidth]{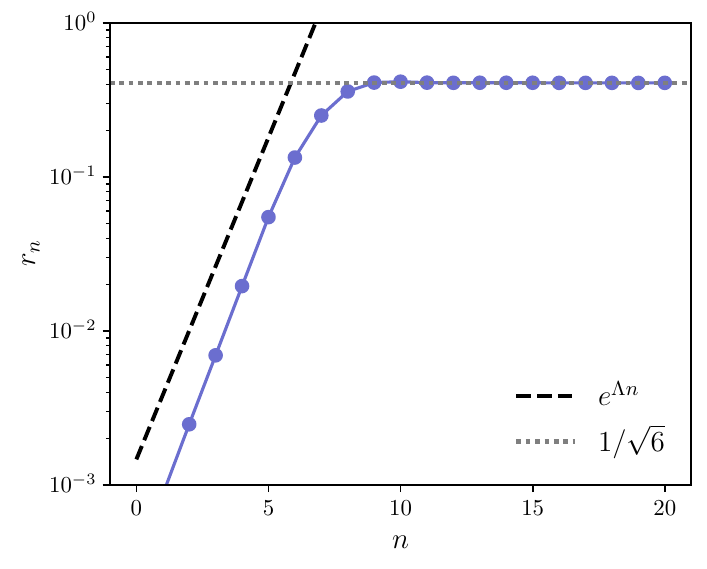}
    \caption{Second moment of the chord function [Eq.~\eqref{eq:chord_r2}] for the Krylov operators
    of the perturbed Arnold's cat map [Eq.~\eqref{eq:q_cat_map}] with $k=0.02$.
    The black dashed line corresponds to exponential growth with the Lyapunov exponent $\Lambda = \ln[(3+\sqrt{5})/2]$.
    The gray horizontal line corresponds to the limit of a uniform chord function.
    The initial operator is given by Eqs.~\eqref{eq:p_representation} and ~\eqref{eq:rho0_torus}
    with $(q_0, p_0) = (0.2, 0.3)$ and $\sigma = 0.45$.
    A Hilbert space dimension of $N=2048$ was used.
}
    \label{fig:r2_chord-Cat}
\end{figure}

Our first result concerns the Lyapunov exponent.
To probe geometric features of the Krylov operators
we use the chord representation~\cite{DEALMEIDA1998265, OZORIODEALMEIDA20061790},
to expand them on the basis of Weyl translations
\begin{equation}
    \hkr_n = \frac{1}{\sqrt{N}} \sum_\alpha \C_n(\alpha) \hat{T}_\alpha
    \label{eq:chord},
\end{equation}
where $\C_n(\alpha) = \Trace(\hat{T}^\dagger_\alpha \hkr_n)/\sqrt{N}$ is the chord function
which has support on the discrete lattice $\alpha = (k/N, l/N)$ with $k,l = 0, \ldots, N-1$.
Geometrically, a chord function concentrated on short chords $\alpha \sim 0$ represents a smooth operator over phase space,
whereas one extending to long chords represents an operator with fine structure.
In this context, stretching is represented as a widening of the chord function's support over increasingly longer chords.
To quantify this, we normalize the distribution as $w_n(\alpha) = \abs{\C_n(\alpha)}^2/\sum_{\beta} \abs{\C_n(\beta)}^2$
and calculate its second moment
\begin{equation}
    r^2_n = \sum_{\alpha} w_n(\alpha) \abs{\xi(\alpha)}^2
    \label{eq:chord_r2},
\end{equation}
where $\xi(\alpha) = (\alpha + 1/2)\mod1 - 1/2$.
The use of $\xi$ takes into account the double-periodicity of the lattice
in the definition of a distance to its origin.
Figure~\ref{fig:r2_chord-Cat} shows that $r_n$ for the Krylov operators increases exponentially at a rate given by the Lyapunov exponent.
The saturation value corresponds to an operator uniformly distributed in the chord representation,
in which case $w_{\text{unif}}(\alpha) = 1/N^2$ and one readily finds that $r_{\text{unif}}^2 = 1/6 + 4/N^2$,
so $r_{\text{unif}} \approx 1/\sqrt{6}$.
We have thus found that the Lyapunov exponent is encoded in the geometry of the Krylov operators.
This is the first result of this Letter.
In the Supplemental Material we provide further numerical evidence for other quantum maps.
In what follows, we provide an analytical derivation of this fact for the unperturbed cat map.

As we will soon see,
chaotic dynamics enter the maximally ergodic regime in Krylov space at an exponential rate,
implying $\hkr_{n+1} \approx \U \hkr_{n}$.
In the case $k=0$, Eq.~\eqref{eq:c_cat_map} is a symplectic linear map
$x' = \M x \mod 1$ with
$\M = \big(\begin{smallmatrix}
  2 & 1\\
  1 & 1
\end{smallmatrix}\big)$
and $x = (q, p)$,
whose quantized version [Eq.~\eqref{eq:q_cat_map}] transforms Weyl translation operators ``classically" as
$\U \hat{T}_{\alpha} =  \hat{T}_{\M^{-1}\alpha}$.
It is then straightforward to calculate the chord function for the $n$-th Krylov operator
$\C_n (\alpha) \approx \C_0(\M^n \alpha)$
by using the orthogonality of the translation operators $\Trace(\hat{T}_\beta^\dagger \hat{T}_\alpha) = N \delta_{\beta\alpha}$.
We then obtain that
\begin{equation}
    r_n^2 \sim
    \sum_{\alpha} \abs{\C_0(\M^n\alpha)}^2 \abs{\xi(\alpha)}^2 =
    \sum_{\alpha} \abs{\C_0(\alpha)}^2 \abs{\xi(\M^{-n}\alpha)}^2
    \label{eq:rn2_1},
\end{equation}
where the last equality holds because the cat map is a discrete torus automorphism~\cite{arnoldi1951principle}.
Now, for a smooth observable, $\C_0(\alpha)$ has support only on short chords $\alpha\sim 0$,
so that, at times earlier than the Ehrenfest time $\tau_E = \ln(N)/\Lambda$
the mod operation does not come into effect,
and thus $\xi(\M^{-n}\alpha) \approx \M^{-n}\alpha$ for $n \leq \tau_E$.
Additionally, diagonalizing $\M$ and factoring out its largest eigenvalue results in
\begin{equation}
    \abs{\xi(\M^{-n}\alpha)}^2 \approx
    e^{2\Lambda n}
    \abs{
        \begin{pmatrix}
            e^{-2\Lambda n} & 0 \\
            0 & 1
        \end{pmatrix}
        U \alpha
    }^2
    \label{eq:xi2_approx},
\end{equation}
where $U$ is the diagonalization matrix.
This rapidly approaches to an exponential growth for $n \geq \Lambda^{-1}$.
Combining these results we finally get
\begin{equation}
    r_n \sim e^{\Lambda n}
    \label{eq:rn_lambda}
\end{equation}
for the early Krylov operators $\hkr_n$ with $\Lambda^{-1} \leq n \leq \tau_E$.
Thus, we have shown analytically that the Lyapunov exponent is encoded in the geometry of the Krylov operators generated before the Ehrenfest time in a paradigmatic example of quantum chaos.

\begin{figure}[t]
    \centering
    \includegraphics[width=\linewidth]{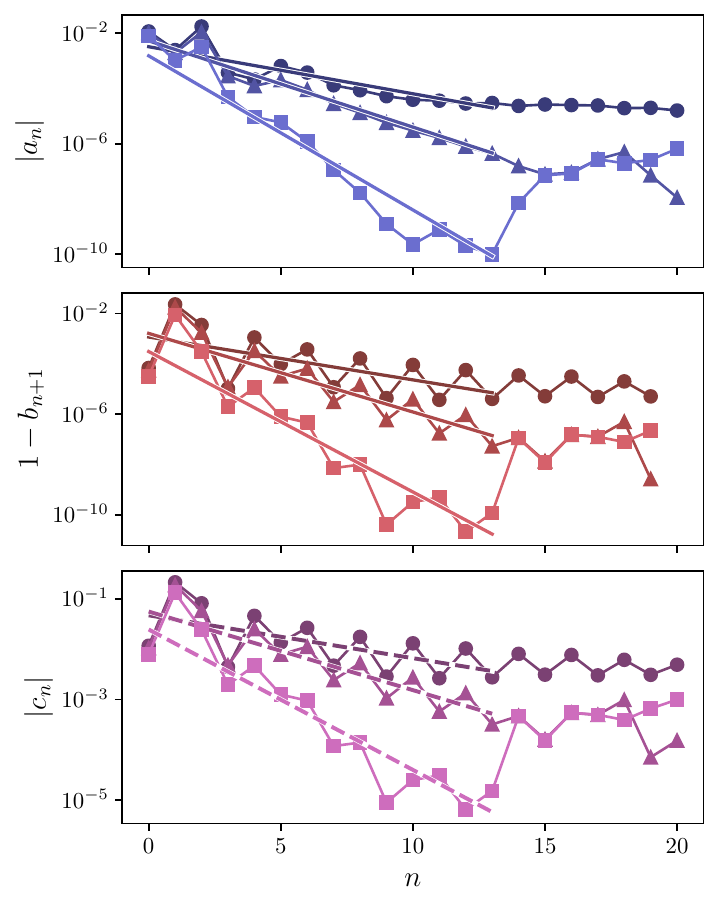}
    \caption{Arnoldi sequences [Eq.~\eqref{eq:arnoldi_sequences}]
    for the perturbed Arnold's cat map [Eq.~\eqref{eq:q_cat_map}] with
    $k=0.02$ (squares), $k=0.025$ (triangles) and $k=0.0275$ (circles).
    The lines are exponential decays with the respective Ruelle-Pollicott resonances
    $\abs{\lambda_1} = 0.526$, $\abs{\lambda_1} = 0.698$ and $\abs{\lambda_1} = 0.822$ (light to dark).
    Solid lines are $\abs{\lambda_1}^{2n}$, and dashed lines $\abs{\lambda_1}^{n}$.
    The initial operator is given by Eqs.~\eqref{eq:p_representation} and ~\eqref{eq:rho0_torus}
    with $(q_0, p_0) = (0.2, 0.3)$ and $\sigma = 0.45$.
    A Hilbert space dimension of $N=2048$ was used.
}
    \label{fig:sequences-Cat}
\end{figure}

We now turn to our results regarding the Ruelle-Pollicott resonances.
In Fig.~\ref{fig:sequences-Cat} we compute the Arnoldi sequences~\eqref{eq:arnoldi_sequences}
of the quantum map~\eqref{eq:q_cat_map} for different degrees of perturbation $k$.
There, we see that the sequences decay exponentially with a rate that depends on $k$.
Fluctuations around the exponential decay may occur due to particularities of the unitary time-evolution.
For example, at $k=0.0275$ (circles) period-2 orbits appear,
inducing persistent oscillations in the non-diagonal sequences.
The variation of their decay rates closely follows that of the Ruelle-Pollicott resonances~\footnote{We have extracted the values $\abs{\lambda_1}$ from Ref.~\cite{PhysRevLett.121.210601}. Note that the values of $k$ reported in Fig. 1 therein may contain a typographic error, as we were able to reproduce those results with the values of $k$ declared here (data not shown).},
such that the Arnoldi sequences follow
$\abs{a_n} \sim \abs{\lambda_1}^{2n}$,
$b_{n+1} \sim 1 - \abs{\lambda_1}^{2(n+1)}$
and
$\abs{c_n} \sim \abs{\lambda_1}^{n}$
in every case.
Thus, the convergence to maximally ergodic Krylov dynamics,
characterized by $a_n \sim 0$, $b_n \sim 1$ and $c_n \sim 0$,
is governed by the Ruelle-Pollicott resonances.
This is the second result of this Letter.
Note the robustness of the results,
since decay of the sequences tracks the dependence of $\abs{\lambda_1}$ on $k$.
In the Supplemental Material, we provide calculations for two additional examples of chaotic maps that further support our observations.
In what follows, we discuss how our findings fit in the broader literature of unitary Krylov dynamics.

Several recent works explore the connection between decay of autocorrelations, Krylov coefficients and Ruelle-Pollicott resonances in chaotic systems.
Ref.~\cite{PhysRevB.111.125103} establishes a direct relation between autocorrelation functions and Krylov coefficients in Floquet dynamics.
In particular, autocorrelations of the form $C(t)=\eta e^{-\delta t}$ give rise to Krylov coefficients that also decay  exponentially, albeit faster.
The faster decay is connected to the fact that in the limit $\eta=1$,
the Krylov coefficients reach the maximally ergodic form after only two time steps~\cite{PhysRevB.111.014309}.
However, the autocorrelations of chaotic maps are not pure exponentials:
they may contain contributions from several Ruelle-Pollicott resonances and oscillatory modes of comparable magnitude.
These deviations prevent the fast convergence to the maximally ergodic regime, so that a decay rate remains visible in the Krylov coefficients.
The converse problem is addressed in Ref.~\cite{duh2026solvablerelaxationdiscreteunitary}.
They propose a simple model where the Verblunsky coefficients are given by $\alpha_n = a\mu^n$,
and find that autocorrelations decay exponentially, but at a rate that in general differs from $\mu$:
more slowly in one phase, which includes the setting of Ref.~\cite{PhysRevB.111.125103}, and faster in the other.
Here, in paradigmatic models of quantum chaos where the Ruelle-Pollicott resonances can be computed independently,
we find that the Verblunsky coefficients decay exponentially with a rate close to the leading nontrivial resonance $\abs{\lambda_1}$ (see the Supplemental Material).

In recent years, Krylov space methods have been established as a powerful framework for characterizing dynamical properties of quantum systems,
with applications ranging from quantum many-body dynamics to black holes and gravity.
Despite this broad applicability,
the connection between Krylov dynamics and properties of classical chaos has remained largely unexplored.
Here, we studied quantized versions of classically chaotic maps,
and have shown that two fundamental signatures of classical chaos are encoded in different structures within Krylov space.
Local instability and stretching, quantified by the Lyapunov exponent, are found in the geometry of the Krylov operators,
as we have demonstrated analytically for a simple but paradigmatic model of quantum chaos.
In contrast, dynamical mixing, characterized by the Ruelle-Pollicott resonances governing the decay of classical correlations,
is reflected in the exponential decay of the Arnoldi sequences
and determines the approach to the maximally ergodic regime in Krylov space.
Our results provide a missing link between the Krylov-space description of quantum dynamics and established signatures of chaos.

\prlsection{Acknowledgments}
This work has been partially supported by CONICET (Grant No.~PIP 11220200100568CO), UBACyT (Grant No.~20020220300049BA).

\bibliography{references}

\clearpage
\onecolumngrid

\begin{center}
  \textbf{\large Supplemental Material for ``\thetitle''}\\[0.5em]
\end{center}

\setcounter{equation}{0}
\setcounter{figure}{0}
\setcounter{table}{0}
\setcounter{page}{1}
\makeatletter
\renewcommand{\theequation}{S\arabic{equation}}
\renewcommand{\thefigure}{S\arabic{figure}}
\renewcommand{\thetable}{S\arabic{table}}

\section*{Results for other maps}
\label{sm:other_maps}
In this section we test the robustness of our results
by studying the Krylov space for two additional examples of chaotic maps.
These are the Harper map,
which is a simple model for the motion of a charged particle under periodic kicks from an external magnetic field~\cite{DANA1995413, Artuso:2011},
and the Chirikov standard map~\cite{Chirikov:2008},
that is derived from the Poincaré section of the kicked rotator.
\begin{figure*}[h]
    \centering
    \includegraphics[width=0.8\linewidth]{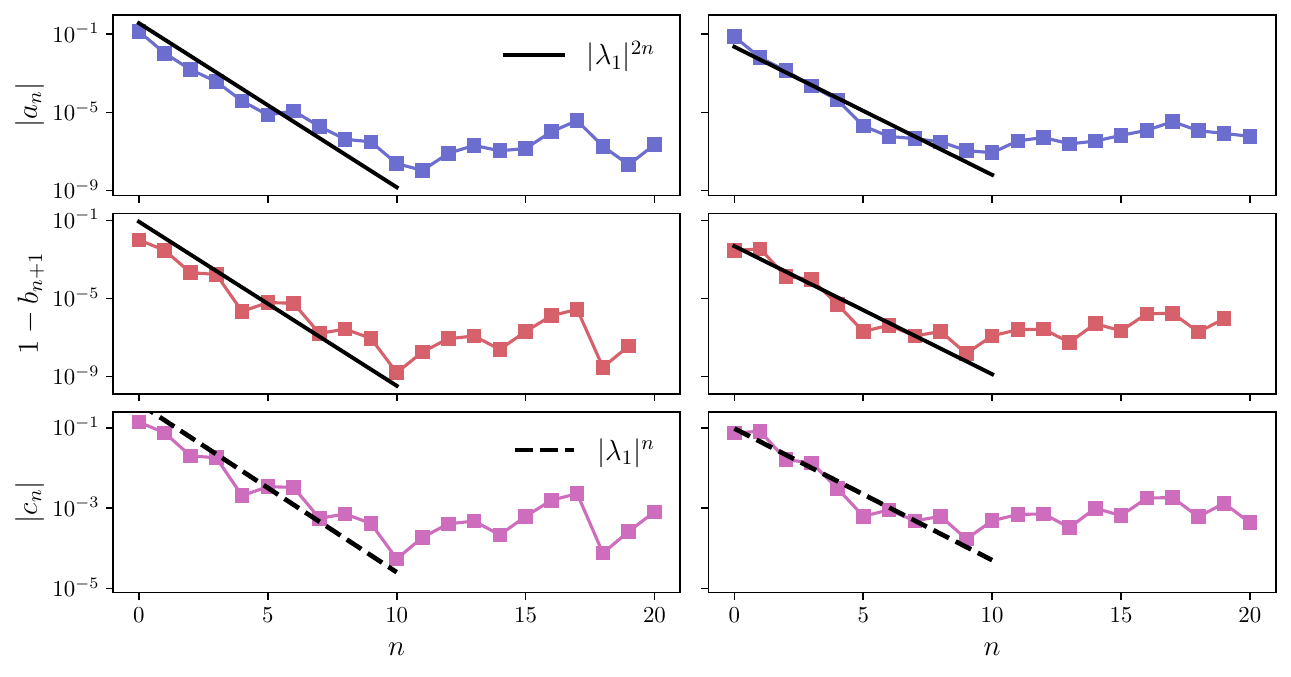}
    \caption{Arnoldi sequences [Eq.~\eqref{eq:arnoldi_sequences}]
        for the Harper map [Eq.~\eqref{eq:q_harper_map}] with $k=0.94$ (left)
    and the Standard map [Eq.~\eqref{eq:q_standard_map}] with $k=19.84$ (right).
    The black lines correspond to exponential decays with the respective Ruelle-Pollicott resonances
    $\abs{\lambda_1} = 0.38$ (left) and $\abs{\lambda_1} = 0.47$ (right).
    The initial operator is given by Eqs.~\eqref{eq:p_representation} and \eqref{eq:rho0_torus} with $(q_0, p_0) = (0.2, 0.3)$ and $\sigma = 0.1$ (left) and $\sigma = 0.45$ (right).
    In both cases a Hilbert space dimension of $N=2048$ was used.
    The behavior of the Arnoldi sequences is equivalent to that of Fig.~\ref{fig:sequences-Cat} on the main text for the cat map.
}
    \label{fig:sequences_two-Harper-Standard}
\end{figure*}

The classical Harper map is defined as
\begin{equation}
    \begin{split}
        p' &= p - k \sin({2\pi q}) \mod 1\\
        q' &= q + k \sin({2\pi p'}) \mod 1
    \label{eq:c_harper_map}.
    \end{split}
\end{equation}
This map generates regular dynamics for $k < 0.11$, while for $k > 0.63$ it is fully chaotic,
without visible regular regions~\cite{PhysRevLett.65.3076}.
Upon quantization, its dynamics is described by the unitary evolution operator
\begin{equation}
    \hat{U} =
    e^{i N k \cos(2\pi \hat{p})}
    e^{i N k\cos(2\pi \hat{q})}
    \label{eq:q_harper_map},
\end{equation}
where $N$ is the dimension of the Hilbert space satisfying the consistency relation $2\pi\hbar N = 1$.
The Chirikov standard map is classically defined through
\begin{equation}
    \begin{split}
        p' &= p + \frac{k}{2\pi} \sin({2\pi q}) \mod 1\\
        q' &= q + p'  \mod 1
    \label{eq:c_standard_map}.
    \end{split}
\end{equation}
For small enough values of $k$, the motion generated by this map is regular
and is bounded in momentum by KAM curves~\cite{GreeneStochastic}.
At a critical value $k_c\approx 0.97$ the last KAM curve breaks,
and there is unbounded diffusion in $p$.
Finally, for large $k$ no significant regular regions remain and the motion is chaotic.
The corresponding evolution operator in the quantum regime is
\begin{equation}
    \hat{U} =
    e^{-i 2\pi N \hat{p}^2}
    e^{i N \frac{k}{2\pi} \cos(2\pi \hat{q})}
    \label{eq:q_standard_map}.
\end{equation}

\begin{figure*}[h]
    \centering
    \includegraphics[width=0.8\linewidth]{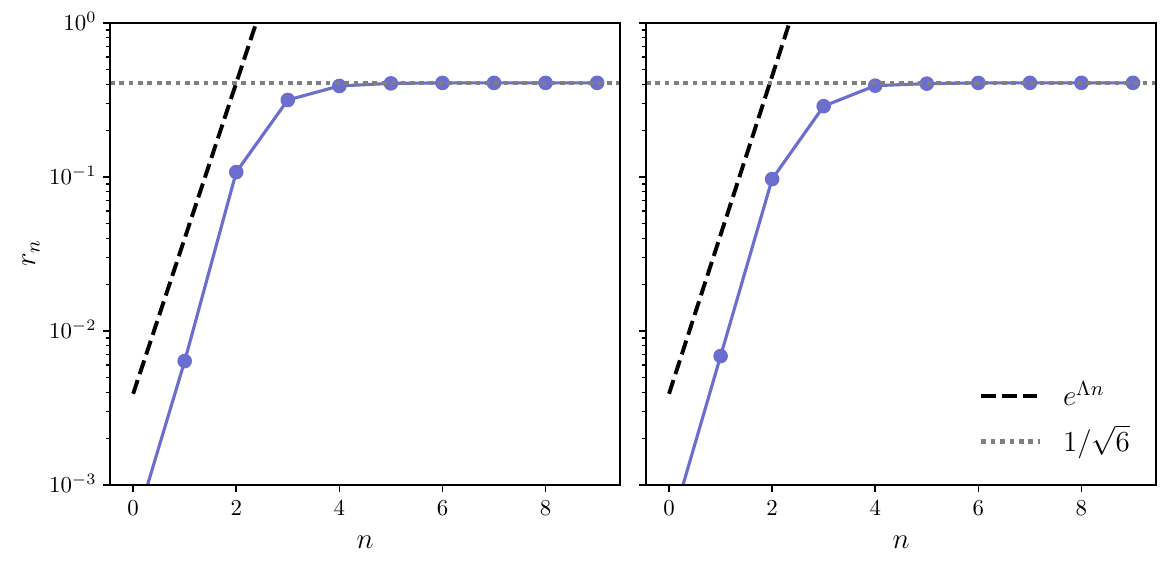}
    \caption{Second moment of the Chord function [Eq.~\eqref{eq:chord_r2}] for the Krylov states
        of the Harper map [Eq.~\eqref{eq:q_harper_map}] with $k=0.94$ (left)
    and the Standard map [Eq.~\eqref{eq:q_standard_map}] with $k=19.84$ (right).
    The black dashed line corresponds to an exponential growth with the respective Lyapunov exponents
    $\Lambda = 2.32$ (left) and $\Lambda = 2.38$ (right).
    The gray horizontal line corresponds to the limit of a uniform Chord function.
    In both cases
    the initial operator is given by Eqs.~\eqref{eq:p_representation} and \eqref{eq:rho0_torus} with $(q_0, p_0) = (0.2, 0.3)$ with $\sigma = 0.45$,
    and a Hilbert space dimension of $N=2048$ was used.
    The behavior of the Chord function is equivalent to that of Fig.~\ref{fig:r2_chord-Cat} on the main text for the cat map.}
    \label{fig:r2_chord_two-Harper-Standard}
\end{figure*}

In Figs.~\ref{fig:sequences_two-Harper-Standard} and \ref{fig:r2_chord_two-Harper-Standard}
we show the results for calculations analogous to that presented in the main text.
These results fully support those shown for the perturbed cat map,
suggesting their robustness.

The unperturbed cat map is a linear map,
and as such its Lyapunov exponent can be obtained simply as $\Lambda = \ln(\alpha)$, where $\alpha$ is the largest eigenvalue of the monodromy matrix.
In nonlinear maps as those present in this section, the Lyapunov exponent is local and depends on the initial condition.
Thus, to compute a physically relevant exponent one should make an average over the points in phase space where the initial distribution is supported.
Here, since we are dealing with smooth observables,
we compute the Lyapunov exponent from the classical evolution
[Eqs.~\eqref{eq:c_harper_map} and \eqref{eq:c_standard_map}] as its average over the whole phase-space.

\section*{Results in terms of Verblunsky coefficients}
\label{sm:verblunsky}
While the Arnoldi iteration has seen ample use
(for example, see Refs.~\cite{PhysRevB.111.014309, PhysRevE.108.054222, PhysRevE.111.L052202, bhattacharya2022operator, PhysRevB.111.174207, PhysRevE.111.014220, Scialchi_2026} to cite a few),
formulations in terms of Verblunsky coefficients have recently been gaining traction~\cite{PhysRevE.111.L052202, PhysRevB.111.125103, jp5f-wpkq, duh2026solvablerelaxationdiscreteunitary}
due to their ubiquity in the mathematics of orthogonal polynomials on the unit circle,
and central role in the Cantero-Moral-Velazquez (CMV) formulation,
allowing a five-diagonal representation of the unitary~\cite{simon2005orthogonal, CANTERO200329, SIMON2007120}.
As mentioned in the main text,
the Arnoldi coefficients are not actually independent,
but can be derived from a single sequence of Verblunsky coefficients.
In this section, we recast our results in such terms.
First, let us discuss the context in which these coefficients arise most naturally,
which is the Szegő recurrence construction of Krylov space.

Since Krylov space is ultimately constructed from the orthonormalization of the system
$\{\U^0 \hrho_0, \U^1 \hrho_0, ...\}$,
the resulting Krylov operators can be expressed as the application of certain polynomials
of the superoperator onto the initial observable
\begin{equation}
    \hkr_n = \sum_{t=0}^{n} \atn{t}{n}\,\U^t \hrho_0 = p_n(\U) \hrho_0
    \label{eq:kr_poly},
\end{equation}
which is why the theory of orthogonal polynomials is so relevant for Krylov constructions.
For unitary $\U$, the relevant class is of those $p_n(z)$ orthogonal on the unit circle $\abs{z} = 1$~\cite{szego, simon2005orthogonal}.
In terms of this construction,
the Krylov operators can be obtained from a dual recursion,
called the Szegő recurrence:
\begin{align}
    b_{n+1} \hkr_{n+1} &= \U \hkr_n - \overline{\alpha_n} \hkr^*_{n}, \label{eq:szego_recurrence_1}\\
    b_{n+1} \hkr^*_{n+1} &= \hkr^*_n - \alpha_n \U \hkr_{n} \label{eq:szego_recurrence_2},
\end{align}
where $b_n$ is the Arnoldi sequence of Eq.~\eqref{eq:arnoldi_sequences},
$\alpha_n$ are the Verblunsky coefficients (which are real for hermitic $\hrho_0$),
and $p_n^*(z) = z^n \overline{p_n(1/\overline{z})}$ is the reciprocal polynomial
($\overline{z}$ denotes complex conjugation).

As previously stated, the Arnoldi sequences can be written in terms of the Verblunsky coefficients.
First, note that the $b_n$ arise as a normalization factor.
By taking the inner product of Eq.~\eqref{eq:szego_recurrence_1} against itself
and using that $\sbraket{\hkr_n^*}{\U \hkr_n} = \overline{\alpha_n}$ one gets
\begin{equation}
    b_{n+1}^2 = 1 - \abs{\alpha_n}^2
    \label{eq:bn_vn}.
\end{equation}
For the other sequences one computes the matrix elements $\U_{m n} = \sbraket{\hkr_m}{\U\hkr_n}$ explicitly.
To do this, take $\U\hkr_n$ from Eq.~\eqref{eq:szego_recurrence_1} and replace into Eq.~\eqref{eq:szego_recurrence_2};
then iterate this recurrence to expand the reciprocal Krylov operators in terms of the usual ones:
\begin{equation}
    \hkr^*_{n+1} = b_{n+1} \hkr^*_n - \alpha_n \hkr_{n+1}
    = - \sum_{l=0}^{n+1} \left(\prod_{k=l}^{n} b_{k+1}\right) \alpha_{l-1} \hkr_l,
    \label{eq:kr*_kr}
\end{equation}
where Eq.~\eqref{eq:bn_vn} was used, and $\alpha_{-1} \equiv -1$.
In the expression above, one should consider the product $\prod_{k=l}^{n} = 1$ for $l\geq n$, .
One can then use this expression in Eq.~\eqref{eq:szego_recurrence_1} and take the inner product with $\hkr_m$ to get
\begin{equation}
    \U_{m n} =
    \begin{cases}
        0 &\text{if}\quad m > n + 1, \\
        b_{n+1} \quad &\text{if}\quad m = n + 1, \\
        - \alpha_{m-1}\overline{\alpha_n} \prod_{k=m}^{n} b_{k+1} &\text{if}\quad m < n + 1.
    \end{cases}
    \label{eq:umn_vn}
\end{equation}
Extracting $a_n = \U_{n n}$ and $c_n = \U_{0 n}$ from the above,
one finally obtains the remaining Arnoldi sequences in terms of the Verblunsky coefficients:
\begin{eqnarray}
    a_n = -\overline{\alpha_n}\alpha_{n-1}, \quad
    c_n = \overline{\alpha_n}\prod_{k=0}^{n} b_{k+1}
    \label{eq:an_cn_vn}.
\end{eqnarray}

\begin{figure*}[h]
    \centering
    \includegraphics[width=0.5\linewidth]{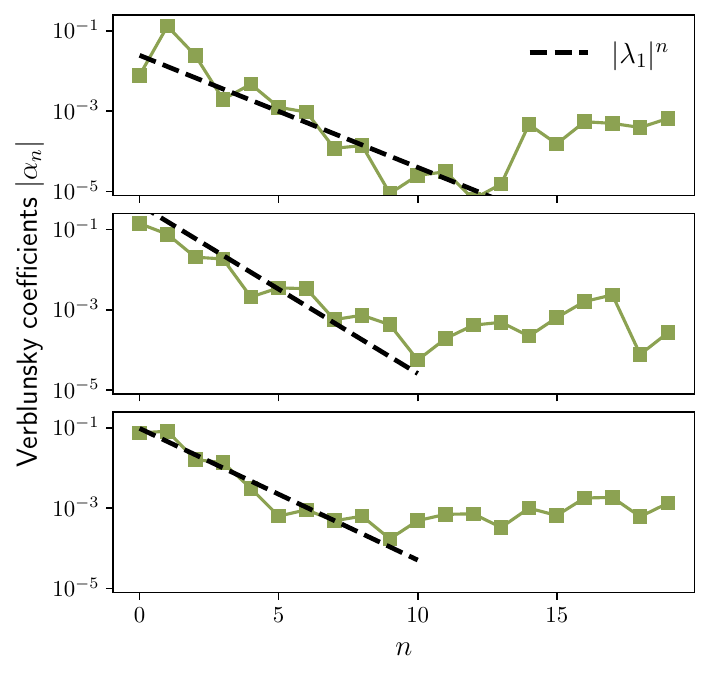}
    \caption{Verblunsky coefficients
    for the perturbed Arnold's cat map [Eq.~\eqref{eq:q_cat_map}] with $k=0.02$ (top),
    for the Harper map [Eq.~\eqref{eq:q_harper_map}] with $k=0.94$ (center)
    and the Standard map [Eq.~\eqref{eq:q_standard_map}] with $k=19.84$ (bottom).
    The black lines correspond to exponential decays with the respective Ruelle-Pollicott resonances
    $\abs{\lambda_1} = 0.526$ (top), $\abs{\lambda_1} = 0.38$ (center) and $\abs{\lambda_1} = 0.47$ (bottom).
    The initial operator is given by Eqs.~\eqref{eq:p_representation} and \eqref{eq:rho0_torus} with $(q_0, p_0) = (0.2, 0.3)$ and $\sigma = 0.1$ for the Harper map and $\sigma = 0.45$ for the cat and Standard maps.
    In all cases a Hilbert space dimension of $N=2048$ was used.}
    \label{fig:verblunsky-all}
\end{figure*}

Figure~\ref{fig:verblunsky-all} shows the Verblunsky coefficients for the maps studied in this manuscript.
We see that the results are analogous and in accordance with those in the main text,
in terms of the Arnoldi sequences:
the Verblunsky coefficients decay exponentially with a rate in accordance to the Ruelle-Pollicott resonances for each map.
We can interpret the different decay rates for the Arnoldi sequences from this result.
Suppose $\abs{\alpha_n} = \alpha \abs{\lambda_1}^n$,
then from Eqs.~\eqref{eq:bn_vn} and \eqref{eq:an_cn_vn} one gets
\begin{align}
    \abs{a_n} &= \alpha^2 \abs{\lambda_1}^{2n -1}, \\
    1 - b_{n+1} &= 1 - \sqrt{1 - \alpha^2 \abs{\lambda_1}^{2n}} = \frac{1}{2} \alpha^2 \abs{\lambda_1}^{2n} + O(\abs{\lambda_1}^{4n}), \\
    \abs{c_n} &= \alpha\abs{\lambda_1}^n \prod_{k=0}^{n} b_{k+1} = \alpha \abs{\lambda_1}^n + O\left(\frac{\alpha^{2(n+1)}}{2^{n+1}}\abs{\lambda_1}^{2n^2}\right)
    \label{eq:an_bn_cn_exp},
\end{align}
which agrees with the decays we see in the Arnoldi sequences throughout the manuscript
[Figs.~\ref{fig:sequences-Cat} and \ref{fig:sequences_two-Harper-Standard}],
that is: $\abs{a_n} \sim 1 - b_{n+1} \sim \abs{\lambda_1}^{2n}$ and $\abs{c_n} \sim \abs{\lambda_1}^n$.

\section*{Chaotic Krylov space of traced observables}
\label{sm:krylov_trace}
In this section we show that it is necessary to consider traceless observables in order for the Ruelle-Pollicott resonances to appear in the Krylov space of chaotic systems.
This is because traced observables have non-zero overlap with the uniform density.
Thus, since chaotic dynamics in a compact phase space relaxes the initial density into a quasi-uniform one,
it will develop a large overlap with the uniform density as it evolves,
leading to a Krylov space completely dominated by it.
We will see that in such case, the Krylov space can be described by a simple model defined by a single parameter depending only on the localization of the initial observable,
drowning out any detail about the system's particular characteristics.

\begin{figure*}[h]
    \centering
    \includegraphics[width=0.8\linewidth]{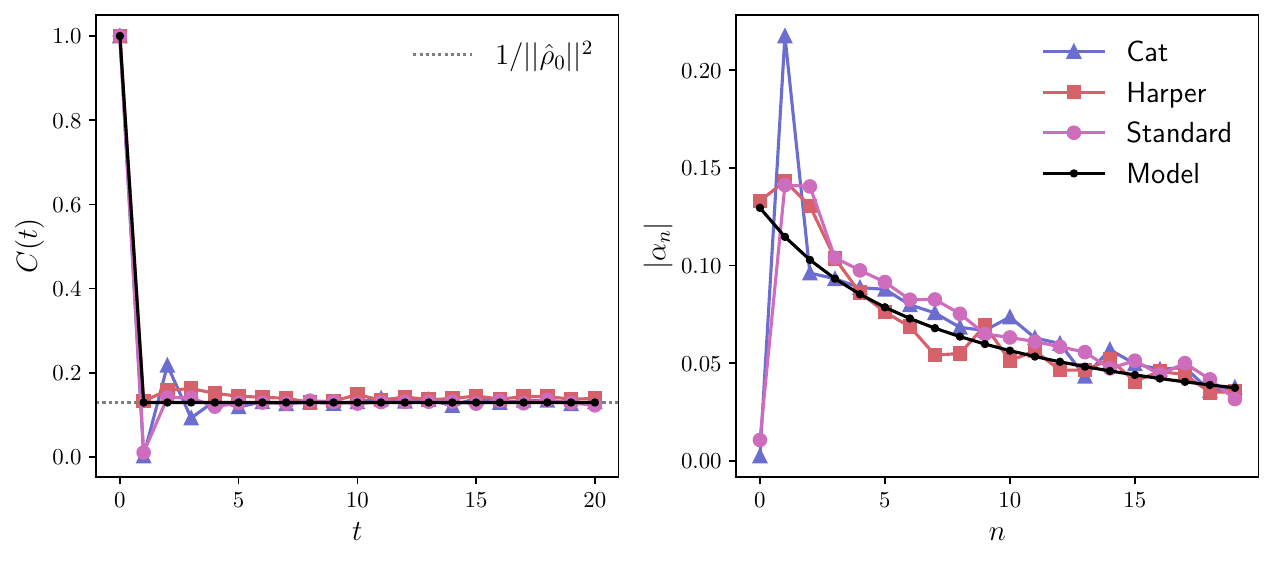}
    \caption{Autocorrelation (left) and Verblunsky coefficients (right)
        for the Cat map [Eq.~\eqref{eq:q_cat_map}] with $k=0.02$ (blue triangles)
        for the Harper map [Eq.~\eqref{eq:q_harper_map}] with $k=0.94$ (red squares)
        and the Standard map [Eq.~\eqref{eq:q_standard_map}] with $k=19.84$ (pink circles).
        The black lines correspond to results from the phenomenological model proposed [Eq.~\eqref{eq:ct_model}].
        In all cases
        the non-traceless initial operator is given by Eqs.~\eqref{eq:p_representation_traceful} and \eqref{eq:rho0_torus_traceful} with $(q_0, p_0) = (0.4, 0.5)$ and $\sigma = 0.1$,
        and a Hilbert space dimension of $N=256$ was used.}
    \label{fig:traceful}
\end{figure*}

Consider an initial observable of the kind considered in the main text,
but without removing the trace:
\begin{equation}
    \hrho_0 = \int dx\,\rho_0(x) \ketbra{\alpha(x)}
    \label{eq:p_representation_traceful},
\end{equation}
where we will still consider $\hrho_0$
defined by a periodicized Gaussian distribution on the unit torus:
\begin{equation}
    \rho_0(q, p) =
    \frac{1}{2\pi\sigma^2} \sum_{n,m\in\mathbb{Z}}
    e^{-\frac{1}{2\sigma^2}\left[(q-q_0-n)^2 + (p-p_0-m)^2\right]}
    \label{eq:rho0_torus_traceful}.
\end{equation}
As can be seen in Fig.~\ref{fig:traceful} (left) the normalized autocorrelation
\begin{equation}
    C(t) = \frac{1}{\norm{\hrho_0}^2}\sbraket{\hrho_0}{\hrho_t}
    \label{eq:ct}
\end{equation}
decays almost instantly to a constant value,
which can be understood from the relaxation of the observable to a quasi-uniform density.
Suppose that after a time $t$ it has relaxed to $\hrho_t \sim \mathbb{I}/N$,
then
\begin{equation}
    C(t) \sim \frac{1}{N \norm{\hrho_0}^2}\sbraket{\hrho_0}{\mathbb{I}}
    = \frac{1}{\norm{\hrho_0}^2} \Trace(\hrho_0) = \frac{1}{\norm{\hrho_0}^2}
    \equiv \gamma
    \label{eq:ct_asymp},
\end{equation}
where we have used the definition of the inner product~\eqref{eq:product} and that $2\pi\hbar N = 1$.
This result coincides with what we see in Fig.~\ref{fig:traceful}.
For an observable of the Gaussian form~\eqref{eq:rho0_torus_traceful} it is easy to calculate
$ \gamma = \norm{\hrho_0}^{-2} = 2\pi(\hbar + 2\sigma^2)$ for $\sigma \ll 1$,
so the asymptotic value is determined by how localized the observable is.

We wish to understand the Krylov space that arises from such dynamics.
Thus, in view of the autocorrelation's behavior,
we propose a phenomenological model defined by an instant decay into a constant value:
\begin{equation}
    C(t) = \delta_{t0}(1 - \gamma) + \gamma
    \label{eq:ct_model},
\end{equation}
and solve the corresponding Krylov space.
A complete and exact solution can be obtained,
which is a fun simple exercise that we will show later.
For now, let us forward the resulting Verblunsky coefficients
\begin{equation}
    \abs{\alpha_n} = \frac{\gamma}{1 + \gamma n}, \quad n\geq 0
    \label{eq:vn_model},
\end{equation}
which as can be seen in Fig.~\ref{fig:traceful} (right) describe very well the overall behavior of the systems considered in this manuscript.
This shows that the influence of the Ruelle-Pollicott resonances argued in the main text is drowned out by the uniform density if one considers a non-traceless initial observable.

Now we solve the Krylov space corresponding to the phenomenological model~\eqref{eq:ct_model}.
Since we start from the autocorrelation,
it is convenient to construct it from the Gram-Schmidt process:
\begin{equation}
    B_n \hkr_n = \hrho_n - \sum_{\nu=0}^{n-1} \sbraket{\hkr_\nu}{\hrho_n} \hkr_\nu
    \label{eq:gram_schmidt},
\end{equation}
with $\hkr_0 = \hrho_0$ (we assume $\hrho_0$ is already normalized)
and $B_n$ are normalization coefficients with $B_0 = 1$.
Start with $n=1$:
\begin{align}
    B_1 \hkr_1 = \hrho_1 - \sbraket{\hkr_0}{\hrho_1}\hkr_0 = \hrho_1 - \gamma \hkr_0, &\quad B_1^2 = 1 - \gamma^2,
    \label{eq:model_sol_1}
\end{align}
since $\sbraket{\hkr_0}{\hrho_1} = \sbraket{\hrho_0}{\hrho_1} = C(1) = \gamma$.
For $n=2$:
\begin{align}
    B_2 \hkr_2 = \hrho_2 - \sbraket{\hkr_0}{\hrho_2}\hkr_0 - \sbraket{\hkr_1}{\hrho_2}\hkr_1 = \hrho_2 - \gamma\hkr_0 - \frac{(1 - \gamma)\gamma}{B_1}\hkr_1, &\quad B_2^2 = 1 - \gamma^2\frac{(1-\gamma)^2}{B_1^2} - \gamma^2
    \label{eq:model_sol_2}.
\end{align}
One may go further to get a feel for the general form of these expressions,
but it will help to express what we already have in a more suggestive manner.
Writing $\gamma = \gamma - 1 + B_0^2$ (since $B_0 = 1$) we can rewrite Eq.~\eqref{eq:model_sol_1} as:
\begin{align}
    B_1 \hkr_1 = \hrho_1 - \frac{\gamma - 1 + B_0^2}{B_0} \hkr_0, &\quad B_1^2 = B_0^2 - \frac{(\gamma - 1 + B_0^2)^2}{B_0^2}
    \label{eq:model_sol_1bis}.
\end{align}
We can do the same with Eq.~\eqref{eq:model_sol_2} by
noting that $1-\gamma^2 = B_1^2$ and so $\gamma(1-\gamma) = \gamma - 1 + B_1^2$,
then
\begin{align}
    B_2 \hkr_2 = \hrho_2 - \frac{\gamma - 1 + B_1^2}{B_1} \hkr_1 - \frac{\gamma - 1 + B_0^2}{B_0} \hkr_0 , &\quad B_2^2 = B_1^2 - \frac{(\gamma - 1 + B_1^2)^2}{B_1^2}
    \label{eq:model_sol_2bis}.
\end{align}
In view of this, one should not doubt of proposing the Ansatz
\begin{equation}
    B_n\hkr_n = \hrho_n - \sum_{\nu=0}^{n-1} \frac{\gamma - 1 +B_\nu^2}{B_\nu}\hkr_\nu
    \label{eq:ansatz1}
\end{equation}
and try to prove it holds via induction.
To do that, it will be useful to expand the Ansatz recursively to get
\begin{equation}
    B_n\hkr_n = \hrho_n - \sum_{\nu=0}^{n-1} \frac{\gamma - 1 +B_\nu^2}{B_\nu^2}
    \left[\prod_{\mu>\nu}^{n-1} \frac{1-\gamma}{B_\mu^2} \right]\hrho_\nu
    \label{eq:ansatz2}.
\end{equation}

We now proceed by induction.
Suppose that exists some $m$ such that for $n\leq m$ Eqs.~\eqref{eq:ansatz1} and \eqref{eq:ansatz2} hold,
then Eq.~\eqref{eq:gram_schmidt} gives the Krylov operator for $m+1$.
We will need the inner products $\sbraket{\hkr_\nu}{\hrho_{m+1}}$,
which from the expression above are
\begin{align}
    \sbraket{\hkr_\nu}{\hrho_{m+1}} = B_\nu^{-1}\left[\gamma - \gamma \sum_{\mu=0}^{\nu-1} b_{\mu, \nu-1} \right] = \frac{\gamma - 1 + B_\nu^2}{B_\nu}, &\quad
      b_{\nu, m} \equiv \frac{\gamma - 1 +B_\nu^2}{B_\nu^2}
    \left[\prod_{\mu>\nu}^{m} \frac{1-\gamma}{B_\mu^2} \right],
\end{align}
where we used that $\sbraket{\hrho_\nu}{\hrho_{m+1}} = \sbraket{\hrho_\mu}{\hrho_{m+1}} = \gamma$ for $m\geq \nu,\mu$
and
\begin{equation}
    B_\nu^2 = 1 - \gamma \sum_{\mu=0}^{\nu-1} b_{\mu, \nu-1},
\end{equation}
which can be derived from Eq~\eqref{eq:ansatz2}
by noting that $\sbraket{\hrho_\nu}{\hkr_\nu} = B_\nu$ [See Eq.~\eqref{eq:gram_schmidt}].
So, writing the Gram-Schmidt process for $m+1$
\begin{equation}
    B_{m+1} \hkr_{m+1} = \hrho_{m+1} - \sum_{\nu=0}^{m} \frac{\gamma-1+B_\nu^2}{B_\nu} \hkr_\nu
    \label{eq:induction},
\end{equation}
we have proven that if the Ansatz holds up $\hkr_m$, then it holds also for $\hkr_{m+1}$.
Since we have seen that it holds for $n=0, 1, 2$, it must hold for all $n$.

The system is not yet solved because we do not know $B_n$.
They arise from normalization, such that Eq.~\eqref{eq:ansatz1} requires
\begin{equation}
    B_n^2 = 1 - \sum_{\nu=0}^{n-1} \frac{(\gamma-1+B_\nu^2)^2}{B_\nu^2}
    = \underbrace{1 - \sum_{\nu=0}^{n-2} \frac{(\gamma-1+B_\nu^2)^2}{B_\nu^2}}_{=B_{n-1}^2} - \frac{(\gamma-1+B_{n-1}^2)^2}{B_{n-1}^2}
    = B_{n-1}^2 - \frac{(\gamma-1+B_{n-1}^2)^2}{B_{n-1}^2}.
\end{equation}
By defining $s_n \equiv B_n^2/(1-\gamma)$ this can be written as a simple recursion
\begin{equation}
    s_n = 2 - \frac{1}{s_{n-1}}
    \label{eq:sn_recursion}
\end{equation}
with initial condition $s_0 = 1/(1-\gamma)$.
It can be solved by proposing $s_n = c/(a+bn) + d$,
noting that the initial condition and the $n\to\infty$ limit imply $c=a\gamma/(1-\gamma)$ and $d=1$, respectively,
and then getting $a$ and $b$ needed for Eq.~\eqref{eq:sn_recursion} to hold.
The solution reads
\begin{equation}
    s_n = \frac{\gamma}{1 + \gamma(n -1)} + 1 \quad\text{so that}\quad B_n^2 = (1-\gamma)\left[\frac{\gamma}{1 + \gamma(n -1)} + 1\right].
\end{equation}
This,
together with Eqs.~\eqref{eq:ansatz1} and \eqref{eq:ansatz2},
which give evolution in terms of the Krylov operators (determining the Krylov wavefunction of Eq.~\eqref{eq:beta_evo}) and the Krylov operators in terms of the evolution, respectively,
completes the solution.

Finally, we can obtain the Verblunsky coefficients by noting that,
on the one hand Eq.~\eqref{eq:gram_schmidt} implies $\sbraket{\hrho_n}{\hkr_n} = B_n^{-1}$,
while the Arnoldi iteration [Eq.~\eqref{eq:arnoldi}] $\sbraket{\hrho_n}{\hkr_n} = \prod_{\nu=1}^{n} b_\nu^{-1}$,
and thus $B_n = \prod_{\nu=1}^{n} b_\nu$ for $n\geq 0$.
From this and Eq.~\eqref{eq:bn_vn} follows
\begin{equation}
    b_{n+1}^2 = \left(\frac{B_{n+1}}{B_{n}} \right)^2 = 1 - \abs{\alpha_n}^2
    \quad\text{so that}\quad
    \abs{\alpha_n}^{2} = 1 - \left(\frac{B_{n+1}}{B_{n}} \right)^2 = \left(\frac{\gamma}{1+\gamma n} \right)^2.
\end{equation}
The rest of the Arnoldi sequences can then be obtained from Eqs.~\eqref{eq:an_cn_vn}.


\end{document}